\documentclass{czjphys}         
\usepackage{euscript,amssymb,epsfig}
\def\RR{{\mathbb R}}

\def\CC{{\mathbb C}}

\def\ZZ{{\mathbb Z}}

\newcommand{\fK}{\mathfrak{K}}

\begin{document}
\title{The MHD $\alpha^2-$dynamo, $\ZZ_2-$graded pseudo-Hermiticity,
level crossings and exceptional points of branching
type\,\footnote{Presented at the 2nd International Workshop
"Pseudo-Hermitian Hamiltonians in Quantum Physics", Prague, Czech
Republic, June 14-16, 2004}}
\authori{Uwe G\"unther, Frank Stefani and Gunter Gerbeth}
\addressi{Research Center Rossendorf, Department of
Magnetohydrodynamics,\\
P.O. Box 510119, D-01314 Dresden, Germany}
\authorii{}
\addressii{}
\authoriii{}    \addressiii{}
\authoriv{}     \addressiv{}
\authorv{}      \addressv{}
\authorvi{}     \addressvi{}
\headauthor{Uwe G\"unther et al.} \headtitle{The MHD
$\alpha^2-$dynamo \ldots} \lastevenhead{Uwe G\"unther et al.: The
MHD $\alpha^2-$dynamo \ldots}
\refnum{}
\daterec{30 July 2004}    
\issuenumber{10}  \year{2004} \setcounter{page}{1075}
\firstpage{1075} \lastpage{1089} 
\maketitle
\begin{abstract}
The spectral branching behavior of the $2\times 2$ operator matrix
of the magneto-hy\-dro\-dy\-na\-mic $\alpha^2-$dynamo is analyzed
numerically. Some qualitative aspects of level crossings are
briefly discussed with the help of a simple toy model which is
based on a $\ZZ_2-$graded-pseudo-Hermitian $2\times 2$ matrix. The
considered issues comprise: the underlying $SU(1,1)$ symmetry and
the Krein space structure of the system, exceptional points of
branching type and diabolic points, as well as the algebraic and
geometric multiplicity of corresponding degenerate eigenvalues.
\end{abstract}

\section{The $2\times 2-$operator matrix of the MHD $\alpha^2-$dynamo}
The magnetic fields of planets, stars and galaxies are maintained
by homogeneous dynamo effects, which can be successfully described
within magnetohydrodynamics (MHD) \cite{krause}. This is achieved
by appropriately combining the Maxwell equations of
electrodynamics with the Navier-Stokes equations of hydrodynamics
--- for certain topologically non-trivial, helical flow fields.
The resulting highly complicated equation systems are subject of
large and cost-intensive computer simulations (see, e.g., Ref.
\cite{nature-1}). Recently, the homogeneous dynamo effect has been
demonstrated in large scale liquid sodium experiments in
Riga/Latvia \cite{riga} and Karlsruhe/Germany \cite{karlsruhe}.
Next generation experiments are currently planned at 7 sites
around the world.

One of the simplest dynamo toy models, which can be regarded as
similar important for MHD dynamo theory like the harmonic
oscillator for quantum mechanics, is the spherically symmetric
$\alpha^2-$dynamo \cite{krause} in its kinematic regime. Its
operator matrix has the form \cite{GS-1}
\be
\hat H_l[\alpha]=\left(\begin{array}{cc}-Q[1] & \alpha\\ Q[\alpha]
& -Q[1]
 \end{array}\right)\label{1}
\ee
and consists of formally selfadjoint blocks
\be
Q[\alpha]:=p\alpha p + \alpha \frac{l(l+1)}{r^2}\label{2}\, .
\ee
($p=-i(\partial_r +1/r)$ is the radial momentum operator.) It
describes the coupled $l-$modes of the poloidal and toroidal
magnetic field components in a mean-field dynamo model with
helical turbulence function ($\alpha-$profile) $\alpha (r)$. The
differential expression (\ref{1}) has the symmetry property
\cite{GS-1}
\be \hat H_l[\alpha]=J\hat H_l^\dagger[\alpha]J ,
\qquad J=\left(\begin{array}{cc}0 & I\\ I & 0
\end{array}\right)\label{3}
\ee
and shows that --- depending on its domain ${\cal D}(\hat
H_l[\alpha])$ and the chosen boundary conditions for the
two-component eigenfunctions $\psi$ --- the corresponding operator
can be pseudo-Hermitian ($J-$selfadjoint). Modulo
 $J-$selfadjoint extensions, this will hold for functions $\psi $ with
 idealized boundary conditions at $r=1$
 \bea
{\cal D}(\hat H_l[\alpha])&:=\left\{ \psi =
\left(\begin{array}{r}\psi_1 \\ \psi_2 \end{array}\right): \
\psi\in \tilde{{\cal H}}\equiv {\cal H} \oplus {\cal H}, \
{\cal H}= L_2(\Omega,r^2 dr),\right. \nonumber\\
& \left. \Omega =[0,1], \ \psi (1)=0, \
\left.r\psi(r)\right|_{r\to 0}\to 0
 \right\},\label{4}
 \eea
whereas the physically realistic boundary conditions
\be
\left.\hat B_l\psi\right|_{r=1} =0, \qquad \hat B_l=\mbox{diag}
[\partial_r+(l+1)/r,1]\label{5}
\ee
lead to a non$-J-$Hermitian operator.

\bfg[htb]                     
\bc                         
\epsfig{file=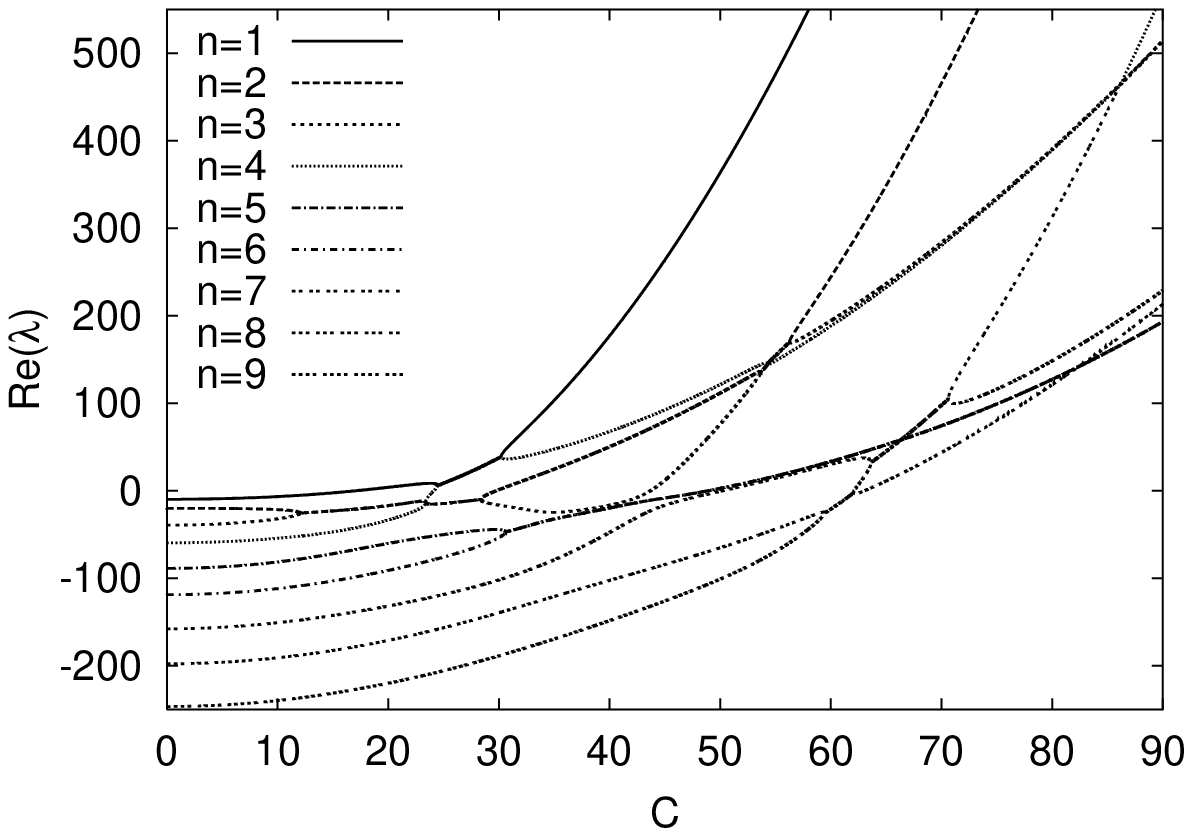,angle=0, width=0.78\textwidth}
\epsfig{file=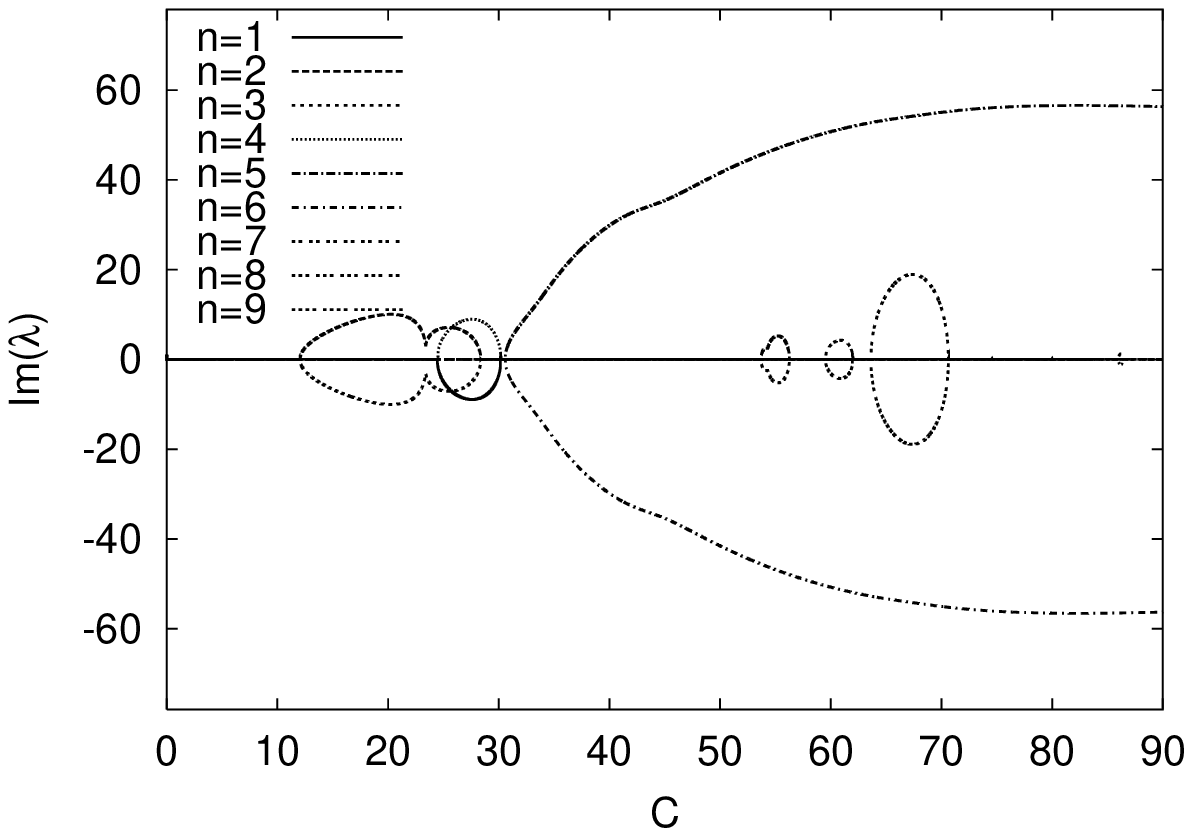,angle=0, width=0.78\textwidth}
\ec                         
\vspace{-2mm} \caption{Large-scale behavior of real and imaginary
parts $\Re (\lambda)$, $\Im (\lambda)$ of the $\alpha^2-$dynamo
spectrum (for physically realistic boundary conditions (\ref{5}))
as function of the scaling parameter $C$ in the concrete
$\alpha-$profile \hfill\mbox{}\newline $\alpha(r)=C\times \left(1-
26.09 \times r^2 +53.64 \times r^3 - 28.22 \times
r^4\right)$.\label{fig1}\vspace{.2cm} \mbox{}}
\efg                       
\bfg[htb]                     
\bc                         
\epsfig{file=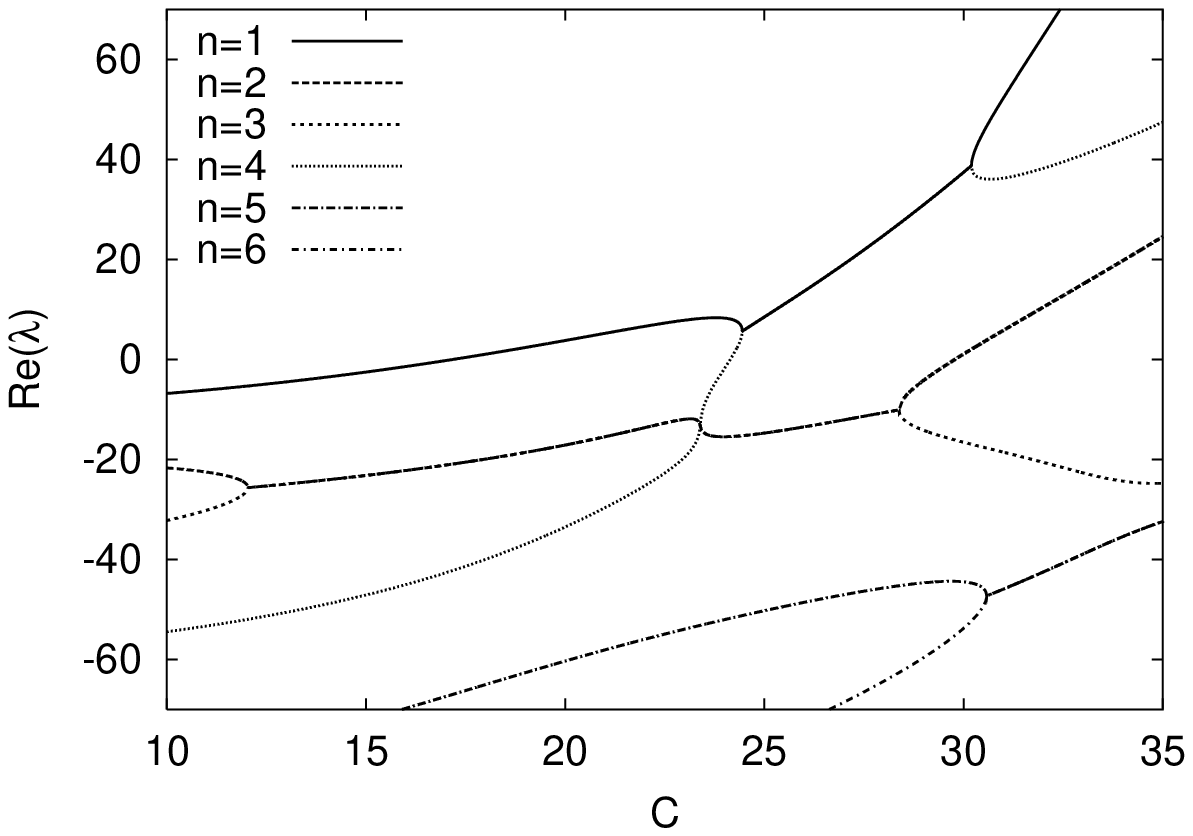,angle=0,
width=0.75\textwidth} \epsfig{file=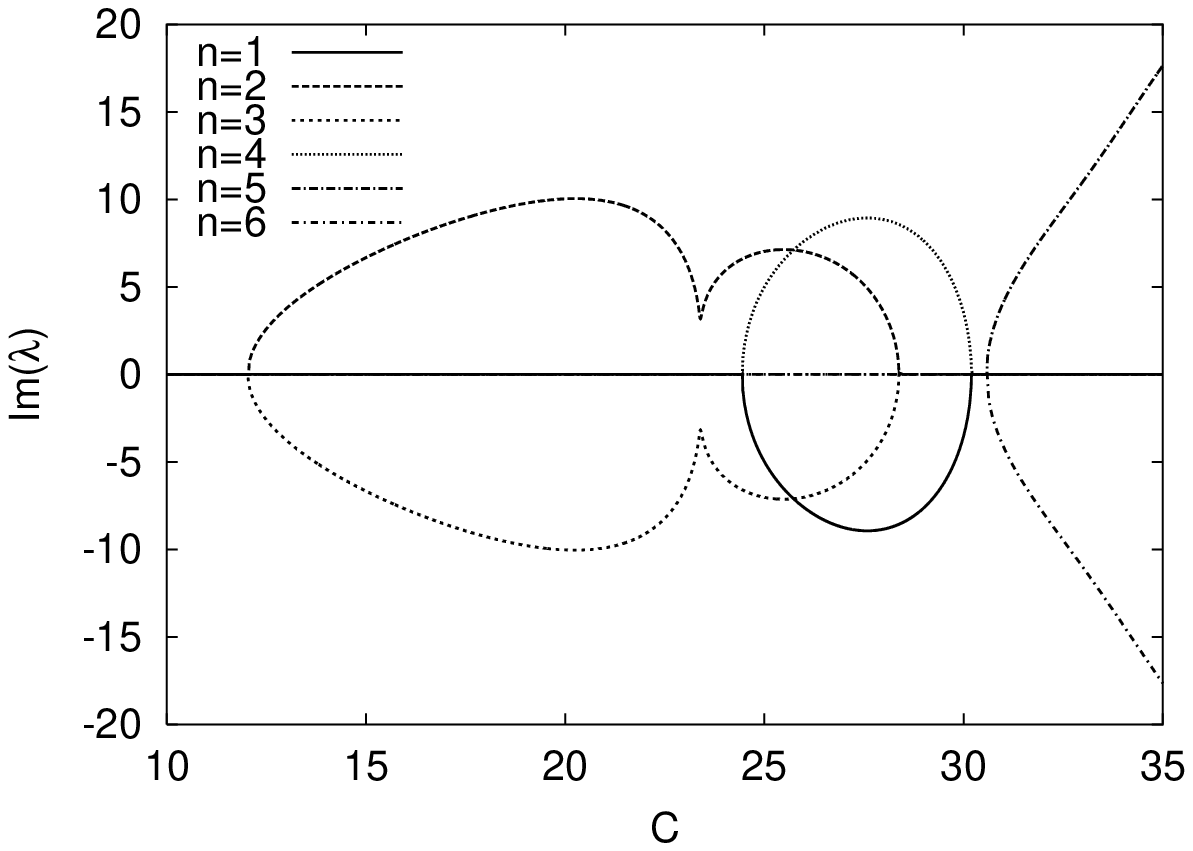,angle=0,
width=0.75\textwidth}
\ec                         
\vspace{-2mm} \caption{Zooming into a scaling region with multiple
level-crossings and branches which "mutually influence" each
other.\label{fig2}\vspace{.2cm} \mbox{}}
\efg                        
\bfg[thb]                     
\bc                         
\epsfig{file=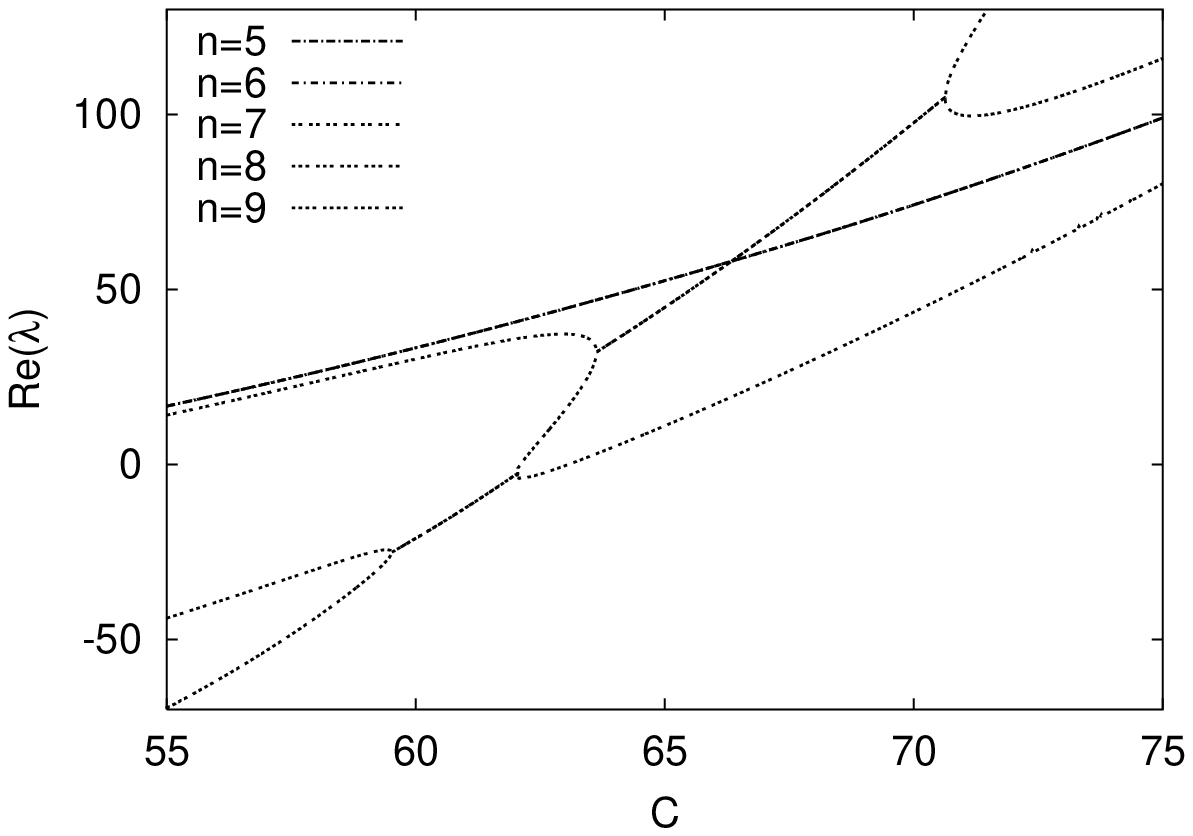,angle=0,
width=0.75\textwidth} \epsfig{file=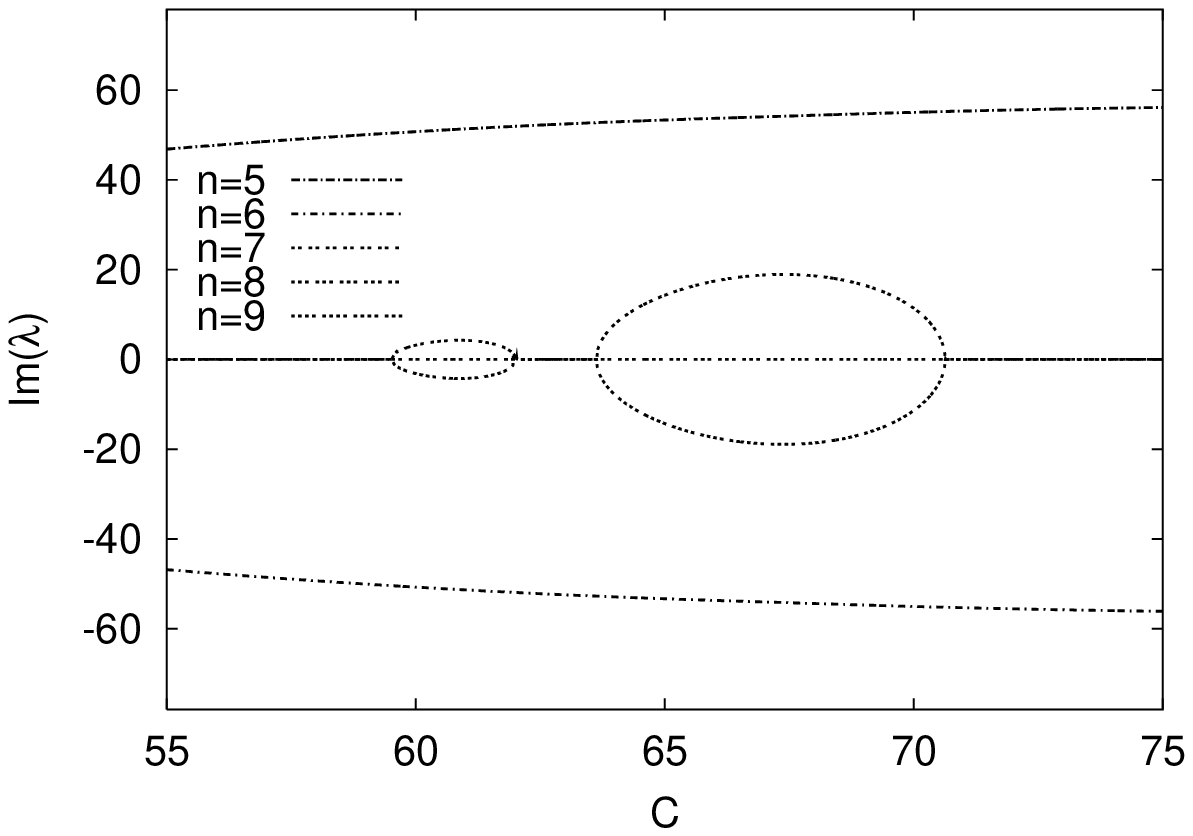,angle=0,
width=0.75\textwidth}
\ec                         
\vspace{-2mm} \caption{Zooming into another scaling region with
multiple level crossings. Interestingly, although locally only two
branches undergo a transition from pairwise real-valued
eigenvalues to pairwise complex-conjugate ones, globally such
transitions occur also for more branches. In the depicted scaling
region three branches participate in mutual transitions. This is a
natural indication of the underlying Riemann surface structure of
the operator spectrum (see, e.g., \cite{heiss-1,marshakov}).
\label{fig3}}
\efg                        
\bfg[thb]                     
\bc                         
\epsfig{file=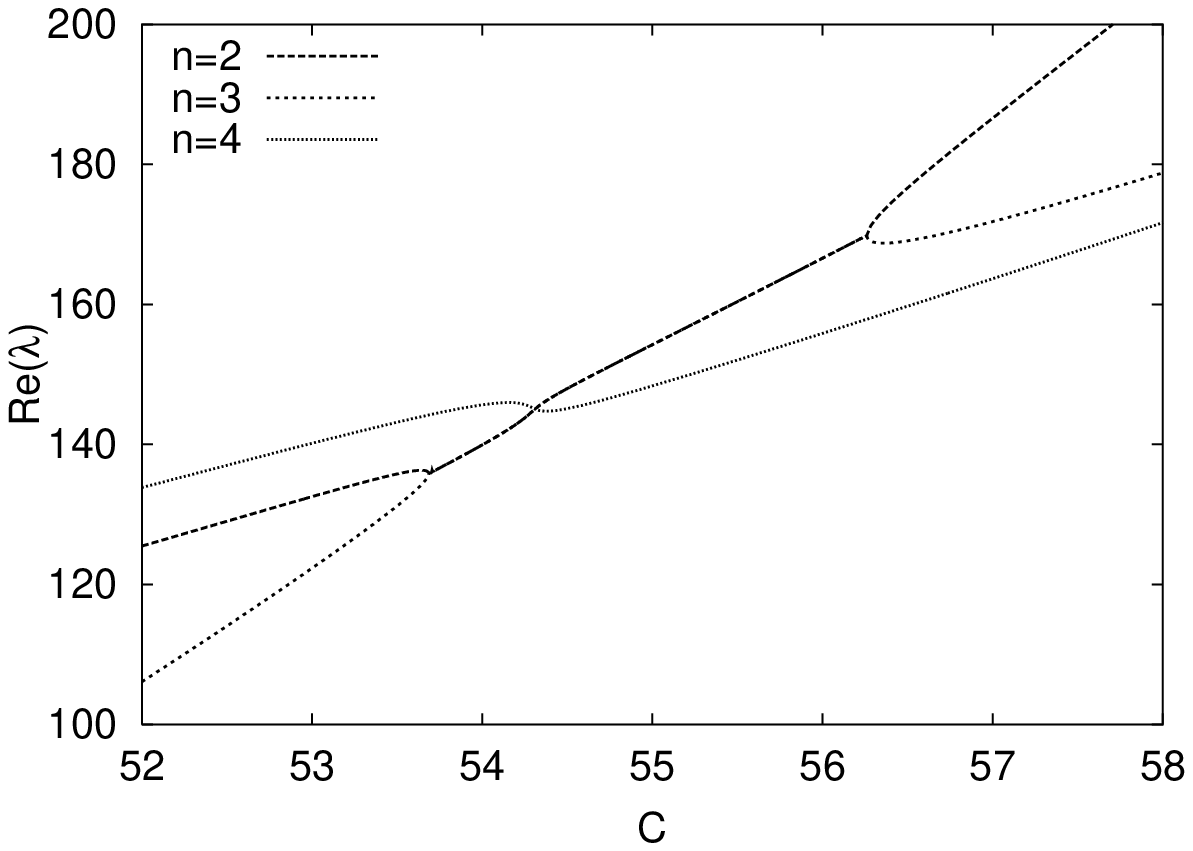,angle=0,
width=0.75\textwidth} \epsfig{file=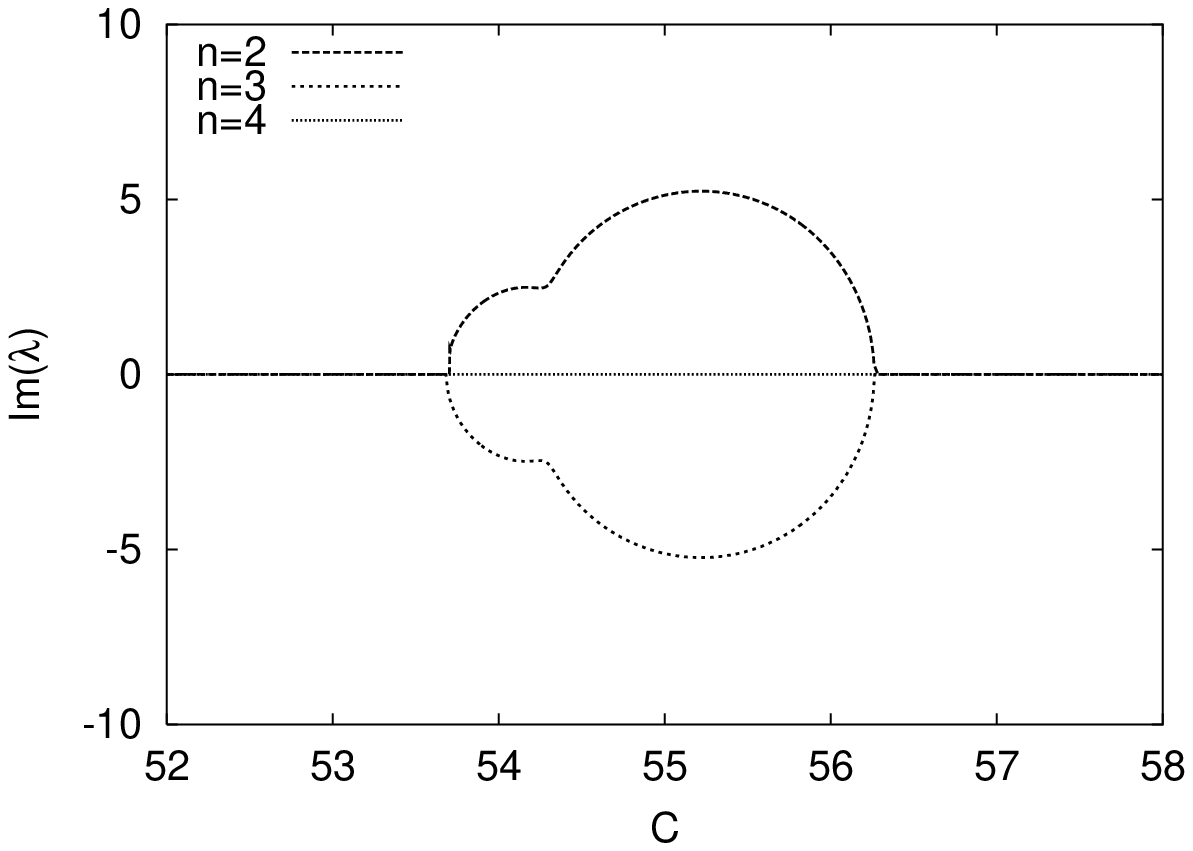,angle=0,
width=0.75\textwidth}
\ec                         
\vspace{-2mm} \caption{The real and imaginary parts of crossing
spectral branches without transition from real to complex
eigenvalues "feel" each other.\label{fig4}\vspace{.2cm} \mbox{}}
\efg                        
\bfg[thb]                     
\bc                         
\epsfig{file=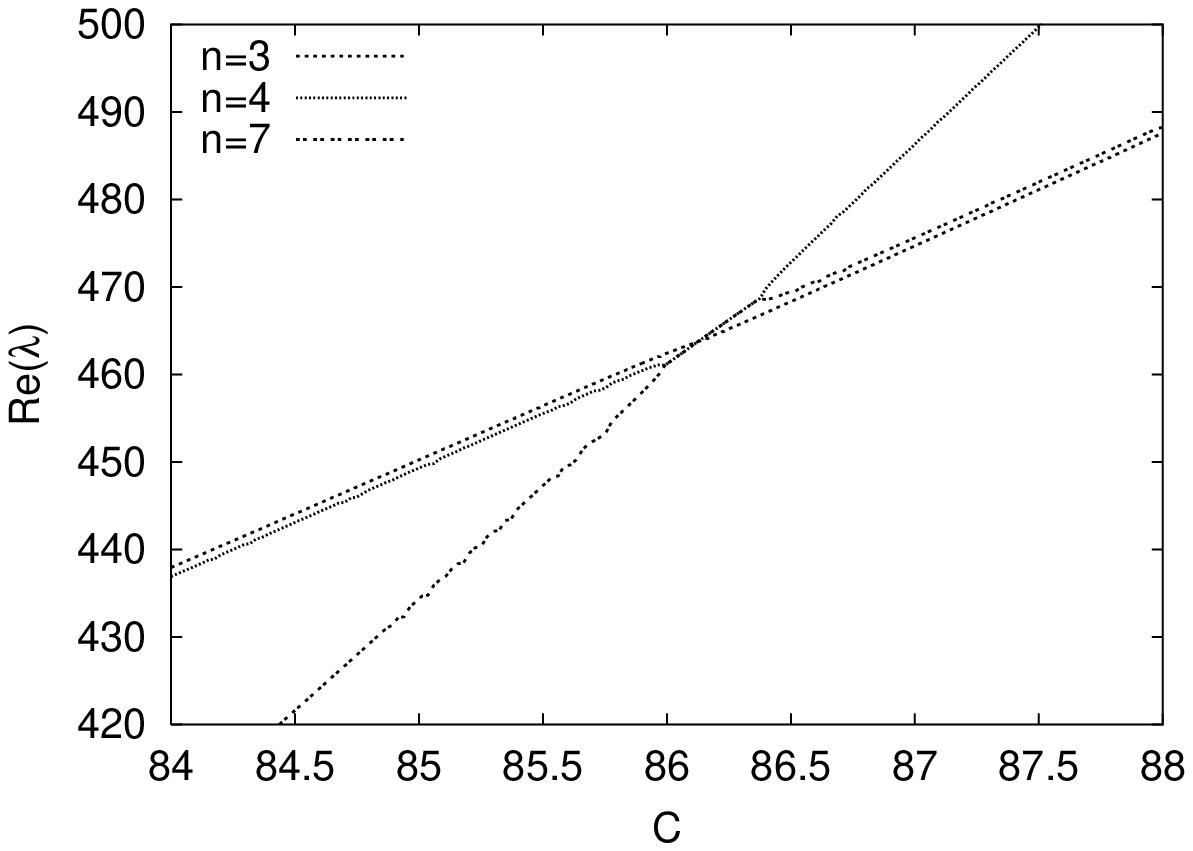,angle=0,
width=0.75\textwidth} \epsfig{file=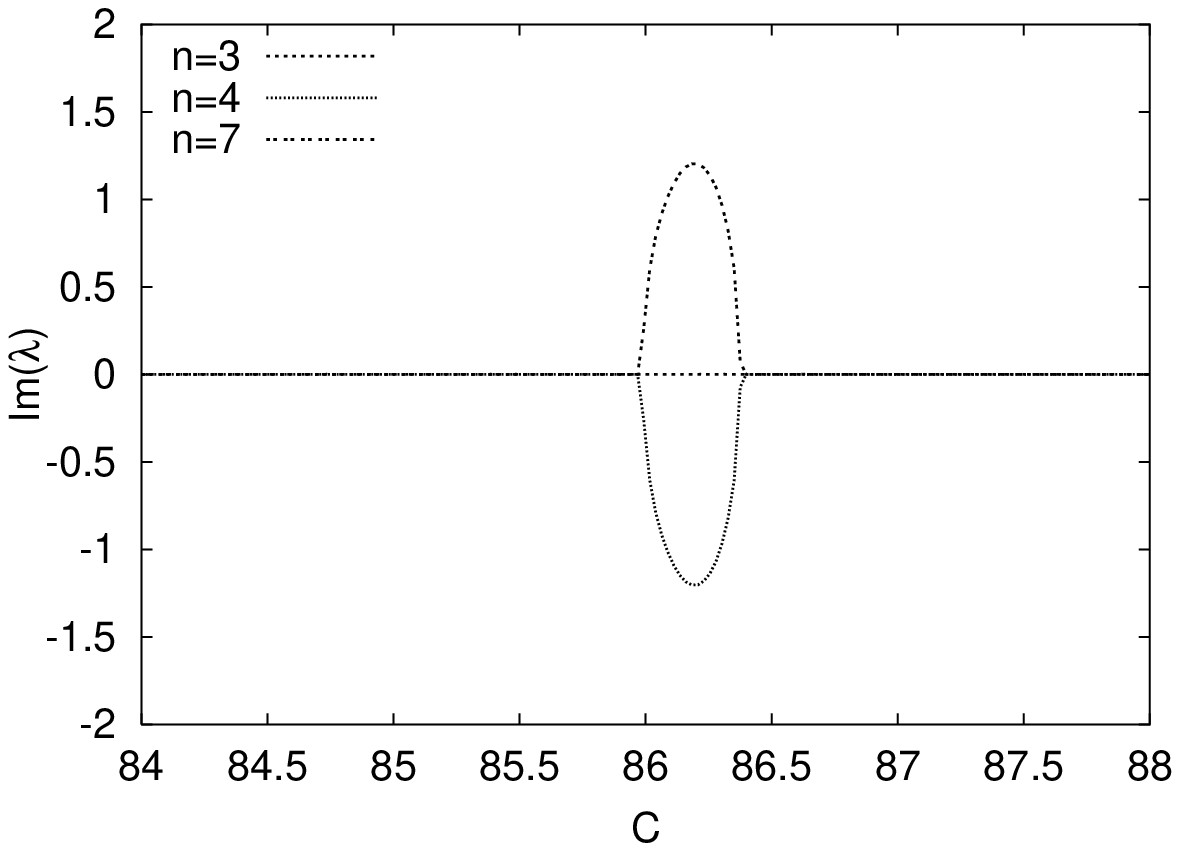,angle=0,
width=0.75\textwidth}
\ec                         
\vspace{-2mm} \caption{The long-term parallel location (repelling)
of different branches without real-to-complex transition is not a
numerical artefact. The branches can be faithfully identified as
different ones.\label{fig5}\vspace{.2cm} \mbox{}} \efg

The dynamo effect starts when the first eigenvalue $\lambda$ of
$\hat H_l[\alpha]$ enters the right half-plane $\Re (\lambda )
>0$. This leads  to an exponential growth $\sim e^{\lambda t}$ of the
solutions of the time dependent kinematic dynamo problem --- and a
corresponding growth of the magnetic field components. In Ref.
\cite{oscil-1} it was demonstrated numerically that this can
happen even on a complex-valued branch of the spectrum, so that
the dynamo can start in an oscillating regime.

In figures \ref{fig1} - \ref{fig5} some typical spectral branches
are shown for an $\alpha^2-$dynamo operator with physically
realistic boundary conditions (\ref{5}). The real and imaginary
components of the eigenvalues for modes with angular mode number
$l=1$ and radial mode numbers $n=1,\ldots,9$ are depicted over a
scaling parameter $C$ of an $\alpha(r)-$profile which is chosen as
quartic polynomial $\alpha(r)=C\times [a_0 + a_2 r^2 + a_3 r^3 +
a_4 r^4]$. Of special physical interest are the critical value
$C_c$, where the spectrum enters the right half-plane $\Re
(\lambda)>0$, as well as the location of those level crossing
points, where two real-valued branches (non-oscillating regime) of
the spectrum meet and continue to evolve as two complex conjugate
branches (oscillating dynamo regime). Within the considered range
of $C$ such transitions occur locally only pairwise --- there are
always only two spectral branches which are locally involved in
such transitions. As shown in fig. \ref{fig3}, globally, there are
more branches involved in mutual transitions. A detailed and
rigorous non-numerical study is still missing. Below, we collect
some few building blocks for such an analysis.

Some first and rough qualitative aspects of the behavior of the
dynamo system at a level crossing point can be easily understood,
e.g., by passing from the eigenvalue problem for the linear pencil
\be \label{6}
\hat L_l[\alpha,\lambda]\psi :=\left(\hat
H_l[\alpha]-\lambda\right)\psi=0
\ee
of the $2\times 2-$operator matrix via substitution
\be
\psi = \left(\begin{array}{r}\psi_1\\ \frac1\alpha [Q(1)+\lambda
]\psi_1
\end{array}\right), \qquad \alpha(r)\neq 0 \label{7}
\ee
to the equivalent eigenvalue problem of the associated quadratic
operator pencil
\bea
L_l[\alpha,\lambda]\psi_1 &\equiv
&\left\{\left[Q[1]+\lambda\right]\frac1\alpha
\left[Q[1]+\lambda\right]-Q[\alpha] \right\}\psi_1 =0 \nonumber \\
&=& (A_2\lambda^2+A_1\lambda+A_0)\psi_1=0.\label{8}
\eea
Solving the quadratic (functional) equation
\be
M_l[\alpha,\lambda]:=(L_l[\alpha,\lambda]\psi_1,\psi_1)=a_2\lambda^2+a_1\lambda
+a_0=0, \qquad a_j:=(A_j\psi_1,\psi_1) \label{9}
\ee
for $\lambda$ we see from the solutions
\be
\lambda_\pm=\frac{1}{2a_2}\left(
-a_1\pm\sqrt{a_1^2-4a_0a_2}\right)\label{10}
\ee
that a level crossing occurs when the discriminant
$\Delta=a_1^2-4a_0a_2$ of Eq. (\ref{9}) vanishes. At the level
crossing points it holds
\be
M_l[\alpha,\lambda]=0,\qquad \partial_\lambda
M_l[\alpha,\lambda]=0 \label{11}
\ee
and the operator pencil has a Jordan-Keldysh chain
\cite{markus,baumg,langer-1} $\{\psi_1,\chi_1,\phi_1\}$,
consisting of the eigenfunction $\psi_1$ and the associated
functions $\chi_1,\phi_1$ which satisfy the relations
\bea
L_l[\alpha,\lambda_0]\psi_1&=&0, \\
L_l[\alpha,\lambda_0]\chi_1+\left.\partial_\lambda
L_l[\alpha,\lambda]\right|_{\lambda_0}\psi_1&=&0,\\
L_l[\alpha,\lambda_0]\phi_1+\left.\partial_\lambda
L_l[\alpha,\lambda]\right|_{\lambda_0}\chi_1+\frac 12
\left.\partial_\lambda^2
L_l[\alpha,\lambda]\right|_{\lambda_0}\psi_1&=&0\label{12}. \eea
Due to the lack of rigorous analytical and non-numerical operator
theoretic studies of the $\alpha^2-$dynamo, we will collect in the
next section some few qualitative facts about a highly simplified
$2\times 2$ matrix model with some rough structural analogies to
the operator matrix (\ref{1}).

\section{$\ZZ_2-$graded pseudo-Hermiticity}
In this section\footnote{The few issues presented in this and the
next section are part of a detailed study given in: U. G\"unther,
{\it $\ZZ_2-$graded pseudo-Hermitian systems: exceptional points,
pseudo-unitary fibrations and nontrivial holonomy}, in
preparation.\label{foot}} we start from the simple model of an
$\alpha^2-$dynamo with idealized boundary conditions (\ref{4}) so
that the corresponding operator $\hat H_l[\alpha]$ is
$J-$selfadjoint ($J-$pseudo-Hermitian) [see relation (\ref{3})].
We will try to get a rough intuitive insight into some of the
basic properties of such a system in the vicinity of a level
crossing point.\footnote{After finishing this proceedings
contribution we became aware that some aspects of level crossings
in $\ZZ_2-$graded pseudo-Hermitian systems had been briefly
discussed earlier --- with the help of different techniques --- in
Refs. \cite{most-2,most-3}. The present analysis overlaps only
marginally with those results and provides a different view on the
subject.}

General pseudo-Hermitian operators $H$ were defined in Refs.
\cite{most-1} as operators which satisfy a relation \be H=\eta
H^\dagger \eta^{-1}.\label{14}
\ee Because of the involution property of $J$, i.e. $J^2=I$,
$J=J^{-1}$, the operator $\hat H_l[\alpha]$ is part of a narrow
subclass of operators with $H=\mu H^\dagger \mu$,
$\mu^2=I,\mu^{-1}=\mu$. To the same subclass belong the
$\cal{PT}-$symmetric Hamiltonians of Refs.
\cite{bender-1,bender-2,bender-3,bender-4}, which are
$\cal{P}-$pseudo-Hermitian and for which necessarily holds
${\cal{P}}^2=I$, as well as the Schr\"odinger Hamiltonian
associated to a Wheeler-DeWitt equation of  minisuperspace quantum
cosmology considered in Ref. \cite{most-1}. A general feature of
systems with involutive pseudo-Hermiticity operators $\mu$ is an
underlying $\ZZ_2-$graded structure of their Hilbert space: Every
state (vector) of the Hilbert space can be naturally split into
(projected onto) $\mu-$even and $\mu-$odd components (for
$\cal{PT}-$symmetric Hamiltonians into parity-even and parity-odd
states, see, e.g., \cite{most-1,znojil}) \be x=P_+ x +P_- x=x_+
+x_-,\qquad P_\pm:=\frac 12 (I\pm \mu)\label{15} \ee with \be \mu
x_\pm=\pm x_\pm.\label{16} \ee Explicitly, this leads to a natural
$\ZZ_2-$grading of the Hilbert space $\tilde{{\cal
H}}\hookrightarrow {\cal H}_+ \oplus {\cal H}_-$ and a Krein space
structure\footnote{For a detailed introduction into the operator
theory over Krein spaces see, e.g., Refs. \cite{langer-1,azizov}.}
\cite{langer-2}: Beside the usual inner product $(.,.)$ of the
Hilbert space $\tilde{{\cal H}}$, which induces a non-negative
norm $(x,x)=(x_+,x_+)+(x_-,x_-)\ge 0$ for $x_\pm \in {\cal H}_\pm
$, one can consider a Krein space $(\fK_\mu,[.,.]_\mu)$ with
indefinite inner product $[x,y]_\mu=(\mu x,y)=(x_+,y_+)-(x_-,y_-)$
(and a corresponding indefinite "norm"). One naturally
distinguishes states (vectors) of positive type $[x,x]_\mu>0$, of
negative type $[x,x]_\mu<0$, and isotropic states $[x,x]_\mu=0$.
(In rough analogy to Minkowski space this corresponds to
time-like, space-like, and light-like vectors.) Because of $\mu
H=H^\dagger \mu$ a $\mu -$selfadjoint ($\mu -$pseudo-Hermitian)
operator is selfadjoint in the Krein space $\fK_\mu $ \be
[Hx,y]_\mu=(\mu Hx,y)=(x,H^\dagger \mu y)=(x,\mu Hy)=[x,Hy]_\mu\,
.\label{17} \ee A natural representation of a $\ZZ_2-$graded
system can be given in terms of 2-component vectors and $2\times
2-$operator matrices \bea &&
x= \left(\begin{array}{c} x_+ \\
x_-
\end{array}\right),  \qquad H= \left(\begin{array}{cc} H_{++} & H_{+-}\\
H_{-+} & H_{--}
\end{array}\right),\nonumber\\ && \mu =\left(\begin{array}{cc} I & 0\\
0 & -I
\end{array}\right),\qquad P_+=\left(\begin{array}{cc} I & 0\\
0 & 0
\end{array}\right),\qquad P_-=\left(\begin{array}{cc} 0 & 0\\
0 & I
\end{array}\right),\label{18}
\eea
where $H_{++}=P_+ HP_+$, etc. The $\mu -$pseudo-Hermiticity
implies  \be H_{++}=H_{++}^\dagger,\qquad
H_{--}=H_{--}^\dagger,\qquad H_{+-}=-H_{-+}^\dagger.\label{19}
\ee

The operator $\hat H_l[\alpha]$ of the $\alpha^2-$dynamo with
idealized boundary conditions can be transformed into this
representation by first diagonalizing the involution operator $J$
\be J\mapsto \mu =S^{-1}JS, \quad
S=\frac{1}{\sqrt{2}}\left(\begin{array}{rr}I&-I\\I&I\end{array}\right).
\ee Applying then the same transformation $S$ to $\hat
H_l[\alpha]$ and the elements of the Hilbert space $\tilde {\cal
H}$ one obtains the equivalent operator \cite{GS-1} \be
\check{H}_l[\alpha] =S^{-1}\hat
H_l[\alpha]S=\frac12\left(\begin{array}{ccc}Q[\alpha -2]+\alpha &
& -Q[\alpha] + \alpha
 \\ Q[\alpha] - \alpha &&Q[-\alpha -2]-\alpha\end{array}\right)
\ee
which acts on 2-vectors
\be
\check{\psi}=\left(\begin{array}{c}\psi_+\\
\psi_-\end{array}\right)
=\frac{1}{\sqrt{2}}\left(\begin{array}{c}\psi_2+\psi_1\\\psi_2-\psi_1\end{array}\right).
\ee
We leave a detailed analysis of this highly non-trivial operator
to future studies.

Instead we try to get a rough qualitative understanding of the
level-crossing in a $\ZZ_2-$graded system. For this purpose, we
consider its simplest example\footnote{This gives a level crossing
with transition from a pair of real-valued eigenvalues to a pair
of complex conjugate eigenvalues. Apart from such level crossing
with real-to-complex transitions, there occur level crossings
without such transitions. They are present in all figs. \ref{fig1}
- \ref{fig5} and can be easily understood in a matrix setup as
crossings of spectral branches which belong to different
$\ZZ_2-$blocks.}
--- the eigenvalue crossing of a $2\times 2$ matrix with the basic
symmetry properties (\ref{19}):
\bea
&& H=\left(\begin{array}{cc} a & b \\
  -b^* & d
\end{array}\right)=\left(\begin{array}{cc} e_0 & 0 \\
  0 & e_0
\end{array}\right)+h,\qquad  h:=\left(\begin{array}{cc} f & b \\
  -b^* & -f
\end{array}\right),\nonumber \\ \nonumber \\&& e_0=(a+d)/2,
\quad f=(a-d)/2,\quad b=b_1+ib_2,\qquad a,d,b_1,b_2\in
\RR.\label{20}
\eea
For the eigenvalues holds
\be
\det\left(H-EI\right)=0 \qquad \Longrightarrow \qquad E=e_0\pm
\sqrt{f^2-b_1^2-b_2^2}\label{21}
\ee
and we see that a level crossing occurs at
\be
\Delta(f,b_1,b_2):=f^2-b_1^2-b_2^2=0.\label{22}
\ee
$\Delta(f,b_1,b_2)=0$ defines a two-dimensional variety (a
double-cone) in the three dimensional parameter space ${\cal M}\ni
(f,b_1,b_2)$ so that for $f\neq 0\neq |b|$ the degeneracy has
co-dimension one. This is in obvious contrast to a Hermitian
$2\times 2$ (spin) matrix problem with
\be
h_s=\left(\begin{array}{cc} f & b \\
  b^* & -f
\end{array}\right)\label{23}
\ee
and eigenvalues
\be
\det\left(H_s-E_sI\right)=0 \qquad \Longrightarrow \qquad
E_s=e_0\pm \sqrt{f^2+b_1^2+b_2^2}.\label{24}
\ee
Here a degeneracy $\Delta_s(f,b_1,b_2):=f^2+b_1^2+b_2^2=0$ occurs
only in the single (diabolic) point $f=b_1=b_2=0$ and the crossing
has co-dimension three \cite{vneumann,berry-1}.

Another difference between the two models is the underlying
symmetry of "iso-energetic" (adiabatic) deformations. Obviously,
the eigenvalues $E$ and $E_s$ in (\ref{21}) and (\ref{24}) are
invariant, respectively, under $SO(1,2)$ and $SO(3)$
transformations in the parameters $f,b_1,b_2$ \cite{novikov}. This
is in natural correspondence with the generators of
"iso-energetic" (adiabatic) transformations which are defined by
the matrices $H$ and $H_s$ themselves. Modulo the Abelean $U(1)$
transformations induced by the elements $ie_0 \left(\begin{array}{cc} 1 & 0 \\
  0 & 1
\end{array}\right)$ these generators are Lie algebra elements of
the type
\bea ih&=&-b_2 \sigma_1 - b_1\sigma_2+if\sigma_3\in
su(1,1)\sim
so(1,2)\sim sl(2,\RR),\label{25}\\
ih_s&=&i\left(b_1 \sigma_1+ b_2\sigma_2+b_3\sigma_3\right)\in
su(2)\sim so(3). \eea ($\sigma_i$ are the Pauli matrices.) Of
course, this interlinks\footnote{Because of its underlying
pseudo-unitary symmetry pseudo-Hermitian (${\cal PT}-$symmetric)
quantum mechanics is not a special variant of quaternionic quantum
mechanics \cite{adler}. One sees this immediately from the
simplest models with $H$ and $H_s$ as Hamiltonians. Quaternionic
structures are naturally connected with $SU(2)$ symmetries
\cite{novikov,adler} and $H_s$, whereas the pseudo-Hermitian model
with Hamiltonian $H$ has invariants which are related to $SU(1,1)$
symmetry transformations. Additionally, we note that $SU(2)$ is a
compact group whereas $SU(1,1)$ is a non-compact one.} also nicely
with the invariance transformations of the two-dimensional Krein
space\footnote{Because of $\dim ({\cal H}_\pm)=1<\infty $ for the
considered $2\times 2$ matrix model, this Krein space is a so
called Pontryagin space \cite{langer-1}.} $(\fK_\mu,[.,.]_\mu)$,
$\mu=\sigma_3$ and the 2-vector Hilbert space ${\cal H}_s$ of the
spin system. The inner product $[.,.]_\mu$ is invariant under
pseudo-unitary  $U(1,1)\sim U(1)\times SU(1,1)$ transformations,
whereas the inner product $(.,.)$ in ${\cal H}_s$ is invariant
under unitary $U(2)\sim U(1)\times SU(2)$ transformations.

Further insight into the structure of the simple pseudo-Hermitian
$2\times 2$ matrix eigenvalue problem can be gained by
diagonalizing the matrix $H$. Before we do this explicitly, we
note that any $2\times 2$ involution matrix $\eta$
\be
\eta^2=I,\qquad \eta=\eta^{-1}=\eta^\dagger\label{26}
\ee
belongs to one of the following classes $\eta_+$ or $\eta_-$
\bea
&\det(\eta_+)=1,\qquad &\eta_+\in\{I,-I\},\label{27}\\
&\det(\eta_-)=-1,\qquad &
\eta_-=a_1\sigma_1+a_2\sigma_2+a_3\sigma_3, \quad
a_1^2+a_2^2+a_3^2=1.\label{28} \eea This means that two involution
matrices $\eta_{1-}$, $\eta_{2-}$ are connected by an $SU(2)\sim
\ZZ_2\times SO(3)$ rotation.

For non-degenerate eigenvalues $E$ the diagonalization is done as
\bea
&& H=SDS^{-1},\qquad D=\left(\begin{array}{cc} e_0-\epsilon & 0 \\
  0 & e_0+\epsilon
\end{array}\right),\qquad S=\left(\begin{array}{cc}\frac{-f+\epsilon}{b^*} \gamma_1 & \frac{-f-\epsilon}{b^*} \gamma_2 \\
  \gamma_1 & \gamma_2
\end{array}\right),\nonumber \\ && \epsilon:=\Delta^{1/2},\quad
\gamma_{1,2}\in\CC.\label{29}
\eea
The complex constants $\gamma_{1,2}$ are still arbitrary and we
can fix them by requiring that the diagonalization preserves the
pseudo-Hermitian structure, i.e. that there exists an involution
matrix $\eta$ so that
\be
D=\eta D^\dagger\eta.\label{30}
\ee
{}From the substitution chain
\bea
H=\mu H^\dagger \mu\ \Longrightarrow\
SDS^{-1}=\mu\left(S^{-1}\right)^\dagger D^\dagger S^\dagger\mu \
\Longrightarrow\  D=S^{-1}\mu\left(S^{-1}\right)^\dagger D^\dagger
S^\dagger\mu S\label{31}
\eea
and relation (\ref{30}) one identifies
\be
\eta =S^\dagger\mu
S=S^{-1}\mu\left(S^{-1}\right)^\dagger=\mu^{-1}=\mu^\dagger\label{32}
\ee
so that
\be
\mu =SS^\dagger \mu SS^\dagger,\qquad \eta =S^\dagger S\eta
S^\dagger S.\label{33}
\ee
{}From the latter equations one concludes that $|\det(S)|^2=1$.
Explicit calculations\footnote{See footnote \ref{foot}.} show that
one can set $\det(S)=1$ and that $S$ is in general neither
unitary, $S^\dagger\neq S^{-1}$, nor pseudo-unitary. Complementary
information can be gained from the explicit structure of the
 matrix $D$. It holds
\bea
\Delta>0:\ \Longrightarrow  &\epsilon=\Delta^{1/2}>0,& \quad D=\left(\begin{array}{cc} e_0-\epsilon & 0 \\
  0 & e_0+\epsilon
\end{array}\right)=D^\dagger\label{34}\\
\Delta<0:\ \Longrightarrow  &\epsilon=i|\epsilon|,& \quad D=\left(\begin{array}{cc} e_0-i|\epsilon| & 0 \\
  0 & e_0+i|\epsilon|
\end{array}\right),\nonumber\\
&& \quad D^\dagger=\left(\begin{array}{cc} e_0+i|\epsilon| & 0 \\
  0 & e_0-i|\epsilon|
\end{array}\right)\label{35}
\eea and one obtains from (\ref{30}) and the explicit form of $S$
in (\ref{29}) (by appropriately tuning the constants
$\gamma_{1,2}$) that
\bea
\Delta>0:&\ f>0:\ & \eta=\left(\begin{array}{cc} -1 & 0 \\
  0 & 1
\end{array}\right),\label{36}\\
&\ f<0:\ & \eta=\left(\begin{array}{cc} 1 & 0 \\
  0 & -1
\end{array}\right),\label{37}\\
\Delta<0:&& \eta=\left(\begin{array}{cc} 0 & 1 \\
  1 & 0
\end{array}\right).\label{38}
\eea We see that smooth changes of the starting
$\mu-$pseudo-Hermitian matrix $H$ lead to qualitative "switchings"
(discontinuities in the mapping  $\mu \mapsto \eta$ which
correspond to rotations in the space of $2\times 2$ involution
matrices $\eta_-$
--- see Eq. (\ref{28})) in the pseudo-Hermitian structure of the
diagonal matrix $D$. These "switchings" occur when the system
intersects critical surfaces in the parameter space: (1) the
surface $f=0$ and (2) the degeneration double cone $\Delta=0$. In
the latter case, the "switching" in the pseudo-Hermiticity
matrices corresponds to a "switching" from a Hermitian diagonal
matrix $D=D^\dagger$ for $\Delta>0$ to a complex-valued diagonal
matrix $D\neq D^\dagger$ for $\Delta<0$. A "switching" occurs also
in the properties of the eigenvectors of the diagonal matrix $D$.
Choosing these eigenvectors in the simplest form as
\bea
D|\pm>=\left(e_0\pm\epsilon\right)|\pm>,& \quad & |+>=\left(\begin{array}{c}  0 \\
  1
\end{array}\right),\quad |->=\left(\begin{array}{c}  1 \\
  0
\end{array}\right)\nonumber\\
<\pm|\pm>=1,&&<+|=(0,1),\quad <-|=(1,0),\label{39}
\eea
one finds for the Krein space inner products
\bea
\Delta>0:&\quad &[\pm,\pm]_\eta=\pm \mbox{sign}(f),\quad
[\pm,\mp]_\eta=0\label{40}\\
\Delta<0:&\quad &[\pm,\pm]_\eta=0,\quad
[\pm,\mp]_\eta=1.\label{41}
\eea
This means that the eigenvectors $|\pm>$ are of positive or
negative type for real-valued eigenvalues, $\Delta>0$, and of
isotropic type for pair-wise complex conjugate eigenvalues,
$\Delta<0$. We explicitly reproduced a basic result of Krein space
theory which is discussed in \cite{azizov} and which is also
implicitly present, e.g., in Refs.
\cite{most-1,bender-3,bender-4}.

\section{Exceptional points}
{}From the structure of $S$ in (\ref{29}) one sees that in the
degeneration limit $\epsilon \to 0$ the determinant $\det (S)$
vanishes, $\det (S)\to 0$, $S$ becomes singular and the
diagonalization (\ref{29}) breaks down. Instead the matrix $D$
turns into a Jordan block. For $\epsilon=0$, $f^2=|b|^2\neq 0$,
$f=\pm |b|$, $a=d\pm 2|b|$ one finds the explicit relation \be
H=SDS^{-1},\qquad D=\left(\begin{array}{cc} E & 1 \\
  0 & E
\end{array}\right),\qquad S=\left(\begin{array}{cc}\mp \frac{|b|}{b^*} & \frac{1}{b^*}  \\
  1 & 0
\end{array}\right), \quad E=d\pm |b|\label{42}.
\ee Additionally, one observes\footnote{The existence of a Jordan block structure
with a single geometric eigenvector is a generic feature of
systems at level-crossing points and was observed for a 1D
pseudo-Hermitian Hamiltonian, e.g., in \cite{dorey}.} that only
$|->$ survives as an (geometric) eigenvector of $D$, whereas $|+>$
is now an associated vector (algebraic eigenvector)
\be (D-EI)|->=0, \qquad (D-EI)|+>=|->, \qquad
(D-EI)^2|+>=0.\label{43}
\ee Hence, the degenerate eigenvalues on the double cone have
algebraic multiplicity two and geometric multiplicity one. The
degeneracy is a branching point degeneracy \cite{berry-2} --- a
double cone of exceptional points of branching type \cite{heiss-2}
in the sense of Kato \cite{kato}.

A singularity of higher order occurs at the center of the double
cone, i.e. in the diabolic point \cite{berry-3} at the origin
$f=|b|=0$ of the parameter space ${\cal M}\ni (f,b_1,b_2)$. There
the transformation matrix $S$ in (\ref{42}) becomes singular and
instead of the Jordan block (\ref{42}) the
matrix $H$ has the form \be H=\left(\begin{array}{cc}E & 0 \\
  0 & E
\end{array}\right), \qquad E=a=d.\ee
This is the same type of co-dimension three degeneracy as in the
case of the spin matrix $H_s$ --- with two (geometric)
eigenvectors, i.e. the algebraic and the geometric multiplicity of
this eigenvalue coincide and equal two.

\bigskip

The question of whether nontrivial holonomy (geometric phases
\cite{berry-1}) and chirality properties of eigenfunctions
\cite{heiss-3} in the vicinity of level-crossing points can play a
role in the dynamics of unstable magnetic field configurations
with field reversals \cite{nature-1} remains as one of the
interesting open issues.

\bigskip
{\small We thank the Professors M.V. Berry, W.D. Heiss, H. Langer,
C. Tretter and M. Znojil for useful comments at different stages
of the work. U.G. acknowledges support from DFG grant GE
682/12-2.}

\end{document}